\documentclass[journal]{IEEEtran}
\usepackage[linesnumbered,ruled]{algorithm2e}
\usepackage{amsmath, amssymb}
\usepackage{graphicx}
\usepackage{subfigure}
\usepackage[noadjust]{cite}
\usepackage{color}

\usepackage{longtable}

\begin{document}
\title{Regions of Attraction Approximation Using Individual Invariance}

\author{Surour~Alaraifi,~\IEEEmembership{Student Member,~IEEE,}
        Seddik~Djouadi,~\IEEEmembership{Member,~IEEE,}
        and~Mohamed~El-Moursi,~\IEEEmembership{Senior~Member,~IEEE}
\thanks{Surour Alaraifi and Mohamed El-Moursi are with the Department
of Electrical Engineering and Computer Science, Masdar Institute of Science and Technology, Khalifa University, Abu Dhabi,
UAE, e-mail: (salaraifi@masdar.ac.ae, melmoursi@masdar.ac.ae).}
\thanks{Seddik Djouadi is with the Department of Electrical Engineering and Computer Science, University of Tennessee,Knoxville, TN 37996-2250, USA, e-mail: (mdjouadi@utk.edu)}
}


\maketitle

\begin{abstract}
Approximating regions of attraction in nonlinear systems require extensive computational and analytical efforts. In this paper, nonlinear vector fields are
recasted as sum of vectors where each individual vector is used to construct an artificial system. The theoretical foundation is provided for a theorem in individual
invariance to relate regions of attraction of artificial systems to the original vector field's region of attraction which leads to significant simplification in
approximating regions of attraction. Several second order examples are used to demonstrate the effectiveness of this theorem.\\
It is also proposed to use this theorem for the transient stability problem in power systems where an algorithm is presented to identify the critical clearing time through sequences of function evaluations. The algorithm is successfully applied on the 3-machine 9-bus system as well as the IEEE 39-bus New England system giving accurate and realistic estimations of the critical clearing time.

\end{abstract}


\IEEEpeerreviewmaketitle

\section{Introduction}
\IEEEPARstart{S}{tability} in nonlinear systems has been and remains a central topic in systems theory. Ever since Lyapunov theory was developed, it played a key role in most of the developments in stability theory afterwards.
The importance of Lyapunov functions is not limited to their ability to directly certify stability of an equilibrium point but extends also to determining stability regions (regions of attraction) \cite{khalil}. \\
One of the first challenges in Lyapunov stability was to characterize a general form of Lyapunov functions for specific vector fields. Hence, many efforts led to the development of somehow general functions for certain types of vector fields \cite{lurepower,14}.\\
A Lure' system is a control system described as a linear system with a sector bounded nonlinearity which allows the use of pre-defined nonlinear Lyapunov functions \cite{lurepower,boyd}. The importance of Lure's systems became more apparent recently with the developments in Semi Definite Programming (SDP) solvers that allow for efficient solutions of linear matrix inequalities (LMI) appearing naturally in Lure-type systems \cite{boyd,boyd2,parrilo}. Thus, Lure-systems remain a topic of interest for researcher in stability analysis and have seen some applications in power systems \cite{lurepower,kostya}.
The literature in stability has seen also attempts to re-define lyapunov functions. In \cite{nonmotonic}, it was shown that lyapunov conditions can be relaxed and Lyapunov function may increase in some subsets as long as it eventually decreases to zero. This relaxation although very helpful and essentially needed, requires extra conditions on higher order derivatives which can limit its applicability is certain cases. Other variations of Lyapunov functions were proposed as well. In \cite{bellman}, the notion of vector Lyapunov functions was introduced where instead of searching for a single function, the search is extended to set of functions giving it more flexibility, however, such results are of practical interest for control design frameworks rather than stability regions' estimation. \cite{wassim,wassim1,wassim2}.\\
There are also other methods to Lyapunov functions such as Zubov's method which can determine the exact region of attraction. Zubov's method is unfortunately theoretical as it requires the solution of a partial differential equation which does not in general has a closed form solution \cite{zbv}.\\
In power systems which represents one of the most advanced and active fields of nonlinear systems' stability analysis \cite{chang1995direct}, 
energy function based approaches dominated the field of stability assessment \cite{paienergy}. 
These approaches rely on the ability to find an appropriate candidate function and on the computation of the critical energy value at the so-called Controlling Unstable Equilibrium Point (UEP) with respect to the pre-defined energy function. This task is involving and difficult and can lead to inaccurate stability assessments if the algorithm deviates slightly from the target point. It is true however that among energy function methods, Controlling UEP-based methods can provide less conservative estimates. The current accepted method for finding the controlling UEP is called the Boundary of stability region based Controlling UEP method (BCU method) \cite{bcu}, \cite{chang1995direct} and was proven to have a high success rate although it was found that the BCU method may fail to predict stability for systems that suffer from instability \cite{thorp,6,paganini}. Nevertheless, the BCU method in power
systems is not designed to estimate regions of attraction, instead it is designed to estimate the relative boundary of stability region with respect to a fault-on system \cite{bcu}.
Recent results in power system stability focused on Lyapunov functions instead but with no significant computational advantages \cite{kostya,milano,sos}.\\
This paper proposes a novel method to estimate regions of attraction of autonomous nonlinear systems. The proposed method overlook the ordinary representation of vector fields and instead reproduce a nonlinear vector field as an algebraic sum of vector fields for which we can develop individual dynamical systems with known regions of attraction, and with the help of the developed theory we are able to reconstruct the region of attraction of the original system from the individual regions. Such approach allows for fast estimation of stability regions and provide an insight on the mathematical interaction of individual vector fields.

\section{Theoretical Background}
Consider an autonomous dynamical system represented by the following differential equation:

\begin{equation}
\dot{x} ̇= f(x), \quad   x\in \Re^{n}
\label{eqn1}
\end{equation}

The solution starting from $x$ at $t=0$ is called the trajectory and is denoted by $\varphi(t,x)$.
$f(x):D\rightarrow\Re^{n}$ is a vector valued function from a domain $D\subset\Re^{n}$ to $\Re^{n}$ that is referred to as the vector field associated with the state vector $x$. It is natural to assume that $f(.)$ satisfies sufficient conditions for uniqueness and existence of solutions. Thus, all required derivatives exist and are continuous.
$x_{e}$ is an equilibrium point if $ f(x_{e})=0 $. An equilibrium point can be either isolated with no other equilibrium point in its vicinity or can be part of a continuum of equilibrium points
(e.g., equilibrium subspace).\\
For the system in (\ref{eqn1}), the region of attraction $\Omega$ of the origin is defined as follows:
\begin{equation}\label{roa}
\Omega =\{x\in D:\lim\limits_{t\rightarrow\infty}\varphi(t,x)= 0\}
\end{equation}

The goal is to achieve the largest possible estimate of $\Omega$. From its definition, $\Omega$ is an invariant set, hence, any trajectory starting in $\Omega$ will remain in it at all time. Generically, the region of attraction of an asymptotically stable equilibrium point is an open, connected and invariant set \cite{khalil}.
These properties are generic and
do not serve in the development of region of attraction estimation algorithms. A very common approach to satisfying these properties is by finding sub-level sets of Lyapunov functions.
For a system defined by (\ref{eqn1}), the region of attraction of the origin  can be estimated by sub-level sets of Lyapunov functions $\Omega_c$ \cite{khalil}:

\[\Omega_c =\{x\in D:V(x)\leq c \} \]
 where $V(x)$ is a Lyapunov function, $c>0$ and  $x_e \in \Omega_c $.
Enlarging such approximation have been the target of extensive research in nonlinear systems. It can be seen however, that the problem of estimating the region of attraction by means of Lyapunov functions is twofold. First, finding the appropriate Lyapunov function, and secondly enlarging the estimated region $\Omega_c$. \\
Lyapunov theory only requires the knowledge of the vector field $f(x)$ (the right hand side of (\ref{eqn1}) and proceeds without any explicit knowledge of solutions, hence, current methods of finding such functions rely on searching for $V(x)$ since the vector field is fixed.\\
In this paper, the vector field is being reproduced as a sum of vector fields such that each individual vector field is positively invariant which leads to a much faster approximation of the region of attraction.

\section{Individual invariance}
In this paper, a new approach is developed to certify stability and estimate the region of attraction for nonlinear system.
This approach rely on converting the nonlinear dynamical system from the form in (\ref{eqn1}) to the following:

\begin{equation}
\dot{x} ̇= f(x) = \sum_{i=1}^{m}f^i(x), \quad   x\in \Re^{n}
\label{msys}
\end{equation}
\\
In (\ref{msys}), if $m=1$ then we arrive at the same definition in (\ref{eqn1}). However, the proposed representation proved to indirectly simplify estimating
the region of attraction as will be given in the following Theorem. \\
\\
\textbf{Theorem 1:\label{Th1}} \textit{ For a system defined by (\ref{msys}), define $\dot{x} ̇^i=f^i (x^i ),\forall i\in[1,m]$. If there exists
$\Omega_e =\{x\in D:\lim\limits_{t\rightarrow\infty}⁡\varphi^i(t,x)=\omega_i, f^i(\omega_i) = 0, \forall i\in[1,m]\}$,
then the set $\Omega_e$ is invariant with respect to the original system (\ref{msys}).}
\\
\\
\textbf{Proof:} From the theorem statement, it is given that every individual artificial system:
\begin{equation}
\dot{x}^i = f^i(x^i) \label{seq}
\end{equation}
is positively invariant in the set $\Omega_e$, hence a trajectory $\varphi^i(t, x)$ initiated at any $x \in \Omega_e$ lies entirely in $\Omega_e$ for $t \geq t_0$.
For $m=1$ we can re-write (\ref{seq}) as an integral equation by applying the fundamental theorem of calculus (see for e.g. \cite{Ross}):
\[x^1(t_1+h) = x_0 + \int_{t_1}^{t_1+h}f^1(x(\tau))d\tau\]
For some positive $h$, since $f^1(\cdot)$ is assumed continuous, the integral above can be approximated using the rectangle method as:
\begin{equation*}
\int_{t_1}^{t_1+h}f^1(x(\tau))d\tau \approx hf^1(x(t_1))
\end{equation*}
for some $t_1>0$. This leads to:
\[x^1(t_1+h) \approx x_0 + hf^1(x(t_1))\]
Let $ x_0 \in \partial\Omega_e$, where $\partial\Omega_e$ denotes the boundary of $\Omega_e$. Then, $x^1(t_1+h)\in\Omega_e$ since $\Omega_e$ is assumed invariant.
A similar observation can be made for all $i\in[1,m]$. For $m=2$ the system becomes:
\[ \dot{\bar{x}} = f^1(\bar{x}) + f^2(\bar{x}) \]
It can similarly be shown that for some positive $h$ we have:
\[\bar{x}(t_2+h) \approx x_0 + hf^1(\bar{x}(t_2)) + hf^2(\bar{x}(t_2))\]
for some $t_2$ such that $0 < t_2 < t_1$ the point $\bar{x}(t_2+h)\in \Omega_e$ since the vector sum $ hf^1(\bar{x}(t_2)) + hf^2(\bar{x}(t_2)) $ will
direct $\bar{x}(t_2+h)$ toward the interior of $\Omega_e$ due to the invariance of the latter. Since this is true for every $ x_0 \in \partial\Omega_e$, then the auxiliary system is positively invariant in the set $\Omega_e$.\\
By induction, for any integer $m \geq 1$ we have:

\[x(t_0+h) \approx x_0 + h\sum_{i=1}^{m}f^i(x(t^m)) \]
\[t_0 \leq t^m \leq t_0+h , \forall i \in [1,m] \]

and $x(t_0+h)\in \Omega_e$ which is a point in the trajectory of system (\ref{msys}), hence, system trajectories cannot leave $\Omega_e$ and that
proves the positive invariance of (\ref{msys}) in the set $\Omega_e$. $\blacksquare$\\

Theorem 1 identifies the region of attraction of an autonomous dynamical system by examining the individual vectors $f^i(x)$ that constitute the vector field $f(x)$. Generally, finding an invariant set for a given vector field is not an easy task, however, theorem 1
does not constraint the choice of individual vectors to a certain form which allows in many cases the construction of simplified individual vector fields with analytically defined invariant sets as will be illustrated in the next section. In such case, the region of attraction is estimated as the intersection of individual invariant sets as in the next corollary.\\

\textbf{Corollary 1:\label{Co1}} \textit{ For a system defined by (\ref{msys}), define $\dot{x} ̇^i=f^i (x^i ),\forall i\in[1,m]$. Let $\Omega_i =\{x\in D:\lim\limits_{t\rightarrow\infty}⁡\varphi^i(t,x)=\omega_i, f^i(\omega_i) = 0\},   \forall i\in[1,m]$, and $\bigcap_{i\in[1,m]}\Omega_i \neq \emptyset  $ then $\Omega_e = \bigcap_{i\in[1,m]}\Omega_i$ is invariant with respect to the original system (\ref{msys}) if $\omega_i \in \Omega_e,\forall i\in[1,m] $.}\\
\\
The proof of this corollary follows immediately from theorem 1. But the importance of this corollary is that it provide a practical guideline to defining $\Omega_e$
with respect to individual systems.\\

\begin{figure}
	\centering
	\includegraphics[clip=true,scale=.7]{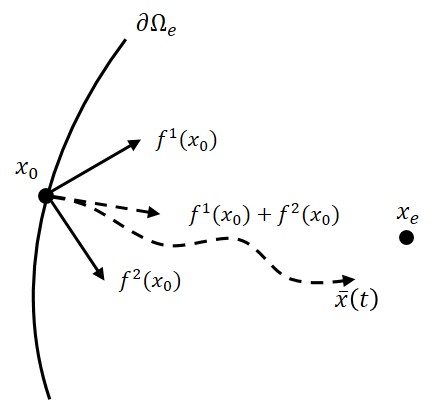}
	\caption{Geometrical interpretation of theorem 1}
	\label{fig1}
\end{figure}

\section{Illustrative examples}

In this section, several examples will be given to illustrate the application of the theorem of individual invariance and its corollary.\\
\\\textit{Example I:}\\

Consider the second order system given by:

\begin{equation} \label{ex1}
\begin{aligned}
&\dot{x}_1 = -b\sin{x_1} -a x_1\\
&\dot{x}_2 = -b\sin{x_2} -a x_2
\end{aligned}
\end{equation}

where $a$ and $b$ are positive real numbers. Let us define the individual vector fields as follows:

\begin{equation*}
\begin{aligned}
f^1(x) = \begin{bmatrix}
-b\sin{x_1}   \\
0 	\end{bmatrix},
f^2(x) = \begin{bmatrix}
0 \\
-b\sin{x_2} \end{bmatrix},
f^3(x) = \begin{bmatrix}
-ax_1 \\
-ax_2 \end{bmatrix}
\end{aligned}
\end{equation*}
\\

Corollary 1 can be applied now by identifying individual invariance sets $\omega_i$ for each artificial system $ \dot{x}^i = f^i(x^i)$.
By inspecting each individual vector, it becomes easy to define each invariant set as follows:

\begin{equation} \label{ex1omega}
\begin{aligned}
&\Omega_1 = \{ x\in \Re^2: |x_1|\leq \pi\}, \\
&\Omega_2 = \{ x\in \Re^2: |x_2|\leq \pi\}, \\
&\Omega_3 = \{ x \in \Re^2 \}
\end{aligned}
\end{equation}

The artificial system defined by $f^1(x)$ has a stable equilibrium subspace $\{0,x_2\}$ and unstable equilibrium subspaces $\{-\pi,x_2\}, \{\pi,x_2\}$ which yield the polyhedron $\Omega_1$ as the invariant set and
a similar argument can be applied to $f^2(x)$. $f^3(x)$ on the other hand represents a linear artificial system with the origin as a globally asymptotically stable equilibrium point. Thus, the whole space is a region of attraction of the origin with respect to the third artificial system.\\
By applying Corollary 1, the intersection of sets in (\ref{ex1omega}) yields the following invariant set:
\[\Omega_e = \{x\in \Re^n: \|x\|_{\infty} \leq \pi \}\]
where  $\|.\|_{\infty}$ is the infinity norm. It is noted that the only equilibrium point in $\Omega_e$ is the origin which is asymptotically stable. Figure 2 illustrates the estimated region of attraction as well as the polyhedrons $\Omega_1$ and $\Omega_2$. \\

\begin{figure}
	\centering
	\includegraphics[clip=true,scale=.5]{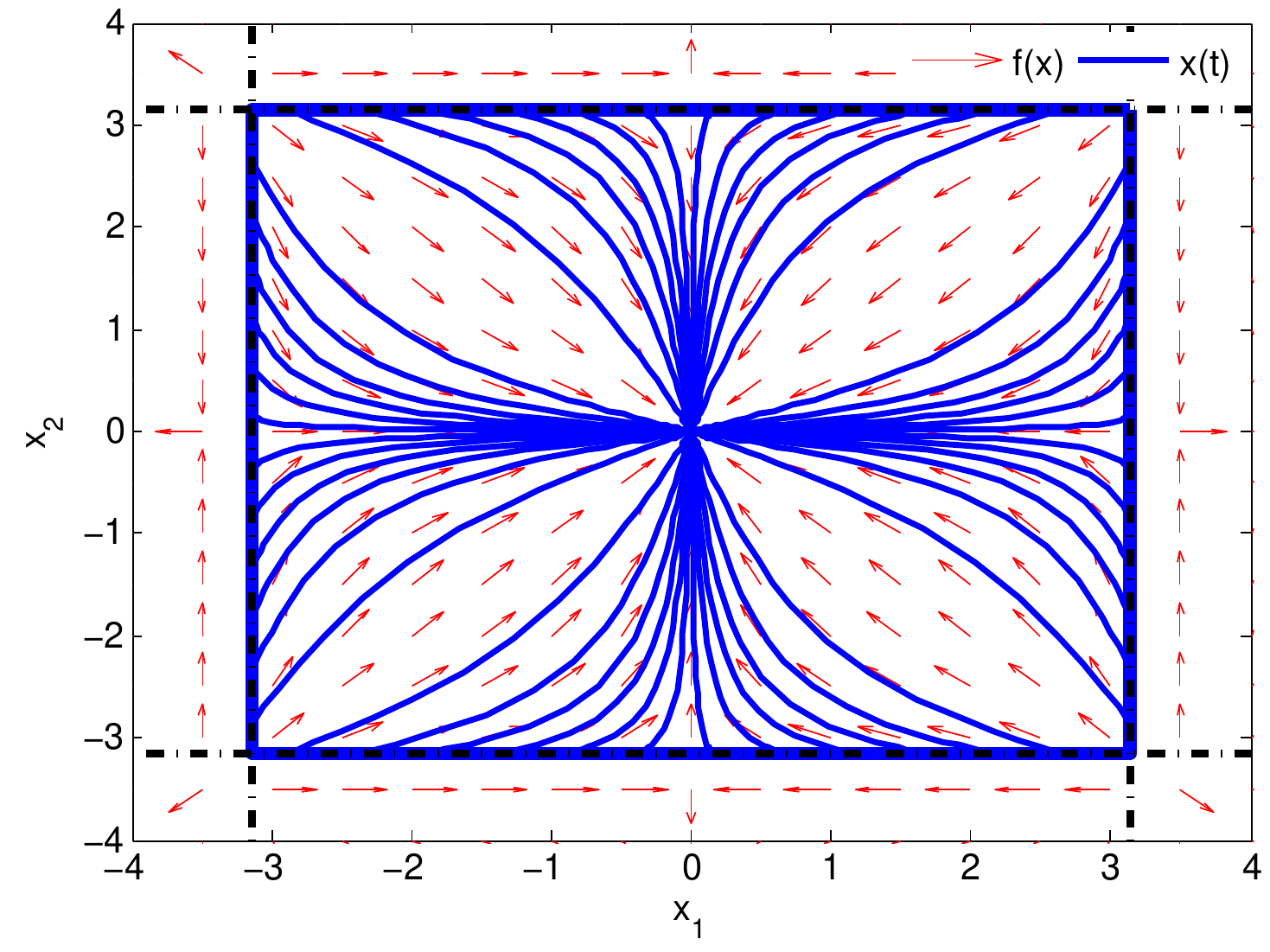}
	\caption{Vector field of (\ref{ex1}) with $a=.1, b=1$}
	\label{fig2}
\end{figure}

\textit{Example II:}\\

Consider the second order system that represents a reduced order two machines system given by \cite{bcu,chang1995direct}:

\begin{equation} \label{ex2}
\begin{aligned}
&\dot{x}_1 = -a_1\sin{x_1} -b \sin{(x_1-x_2)}\\
&\dot{x}_2 = -a_2\sin{x_2} -b \sin{(x_2-x_1)}
\end{aligned}
\end{equation}

where $a_1, a_2$ and $b$ are positive real numbers. Let us define the individual vector fields as follows:

\[
f^1(x) = \begin{bmatrix}
-a_1\sin{x_1}   \\
0 	\end{bmatrix},
f^2(x) = \begin{bmatrix}
0 \\
-a_2\sin{x_2} \end{bmatrix},\]
\[f^3(x) = \begin{bmatrix}
-b \sin{(x_1-x_2)} \\
-b \sin{(x_2-x_1)} \end{bmatrix}
\]

Similar to example I, Corollary 1 can be applied now by identifying individual invariance sets $\omega_i$ for each artificial system $ \dot{x}^i = f^i(x^i)$ as follows:

\begin{equation} \label{ex2omega}
\begin{aligned}
&\Omega_1 = \{ x\in \Re^2: |x_1|\leq \pi\}, \\
&\Omega_2 = \{ x\in \Re^2: |x_2|\leq \pi\}, \\
&\Omega_3 = \{ x\in \Re^2: |x_1 - x_2|\leq \pi \}
\end{aligned}
\end{equation}

The artificial systems 1 and 2 are the exact systems in the previous example with similar invariant sets.$f^3(x)$ represents a coupled nonlinear interaction
with a stable equilibrium subspace $\{x\in\Re^2: x_1-x_2 = 0 \}$ and unstable equilibrium subspaces at
$\{x\in\Re^2: x_1-x_2 = -\pi \}$ and $\{x\in\Re^2: x_1-x_2 = \pi \}$. Thus, the polyhedron defined by $ \Omega_3$ is an invariant set.
\\
By applying Corollary 1, the intersection of sets in (\ref{ex2omega}) yields the following invariant set:
\[\Omega_e = \{x\in \Re^n: \|x\|_{\infty} \leq \pi,|x_1 - x_2|\leq \pi  \}\]

Figure 3 illustrates the estimated region of attraction as well as the polyhedron $\Omega_1, \Omega_2$ and $\Omega_3$. \\

\begin{figure}
	\centering
	\includegraphics[clip=true,scale=.5]{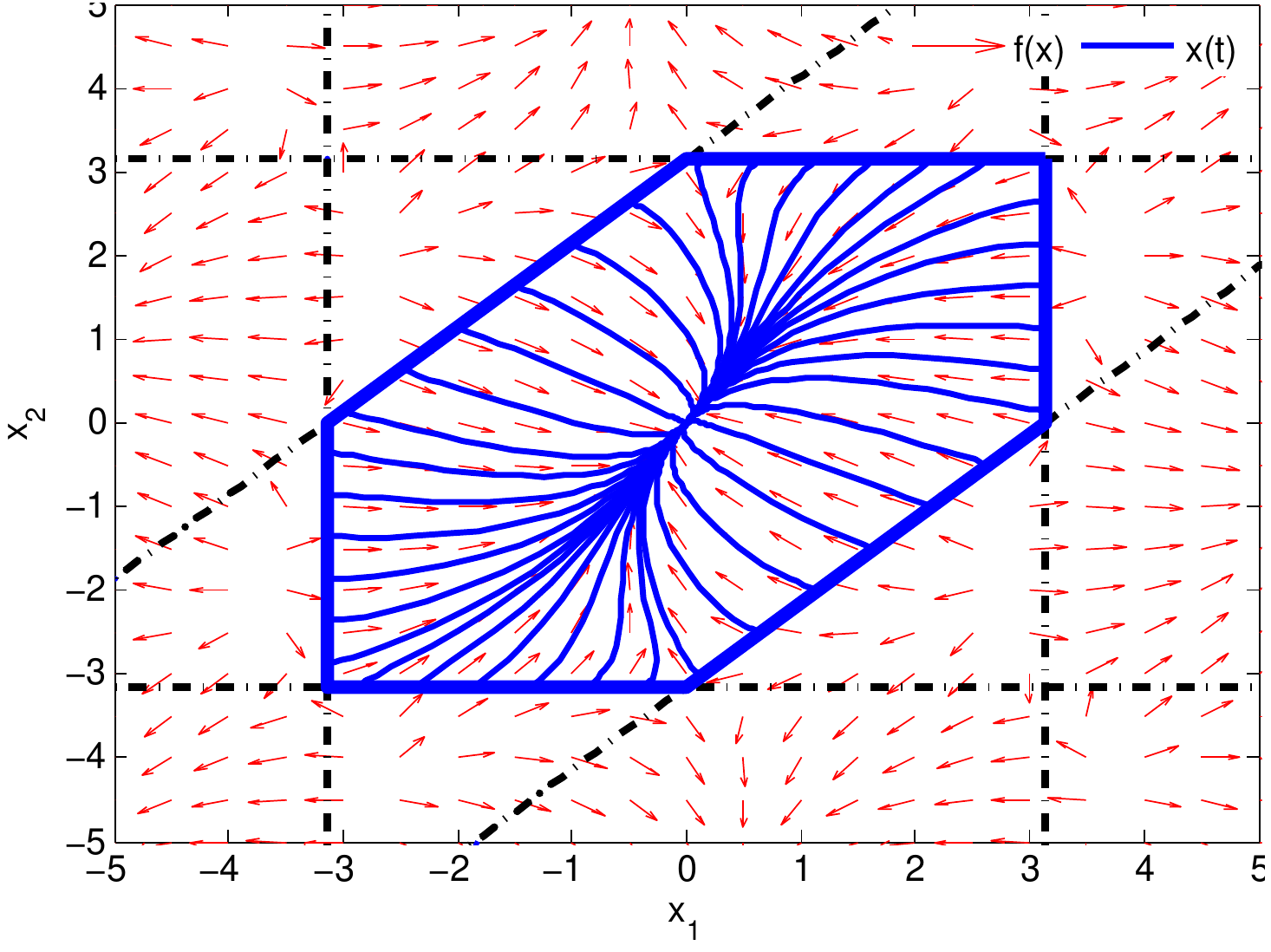}
	\caption{Vector field of (\ref{ex2}) with $a_1=1, a_2=0.5, b=0.5$}
	\label{fig3}
\end{figure}

\textit{Example III:}\\

Consider the second order system that represents a reduced order flux decay model given by \cite{paienergy}:

\begin{equation} \label{ex3}
\begin{aligned}
&\dot{x}_1 = -ax_2\sin{x_1}\\
&\dot{x}_2 = -bx_2 + c\cos{x_1}
\end{aligned}
\end{equation}

where $a, b$ and $c$ are positive real numbers. Let us define the individual vector fields as follows:

\[
f^1(x) = \begin{bmatrix}
-ax_2\sin{x_1}   \\
0 	\end{bmatrix},
f^2(x) = \begin{bmatrix}
0 \\
-bx_2 + c\cos{x_1} \end{bmatrix}\]

Similar to example I, Corollary 1 can be applied now by identifying individual invariance sets $\omega_i$ for each artificial system $ \dot{x}^i = f^i(x^i)$ as follows:

\begin{equation} \label{ex3omega}
\begin{aligned}
&\Omega_1 = \{ x\in \Re^2: |x_1|\leq \pi, x_2 \geq 0\}, \\
&\Omega_2 = \{ x\in \Re^2: |x_1|\leq \frac{\pi}{2} \}, \\
\end{aligned}
\end{equation}

By inspection, the artificial system 1 requires the non-negativity of $x_2$ in order to maintain the negativity of the quantity $-ax_2\sin{x_1}$ and it also requires $x_1$ to be in the set $\{ x_1: |x_1|\leq \pi\}$ which guarantees that trajectories of the artificial system will converge to the continuum of stable equilibrium points defined as $\{x\in\Re^2: x_1 = 0, x_2 \geq 0 \}$. $f^2(x)$ represents a more complicated behavior, with continuum of stable equilibrium points defined as $\{x\in\Re^2: x_2 = \frac{c}{b}\cos{x_1} \}$. However, since artificial system 1 restricts $x_2$ to be nonnegative, then, it is easy to see that $x_1$ lies in the set $\Omega_2$ as defined above.\\

By applying Corollary 1, the intersection of sets in (\ref{ex3omega}) yields the following invariant set:

\[\Omega_e = \{x\in \Re^n: |x_1|\leq \frac{\pi}{2},x_2\geq 0  \}\]

The equilibrium point in $\Omega_e$ is $x_e=\{0,\frac{c}{b}\}$ and is unique and asymptotically stable. Figure 4 illustrates the estimated region of attraction as well as the polytope $\Omega_e$.\\

\begin{figure}
	\centering
	\includegraphics[clip=true,scale=.5]{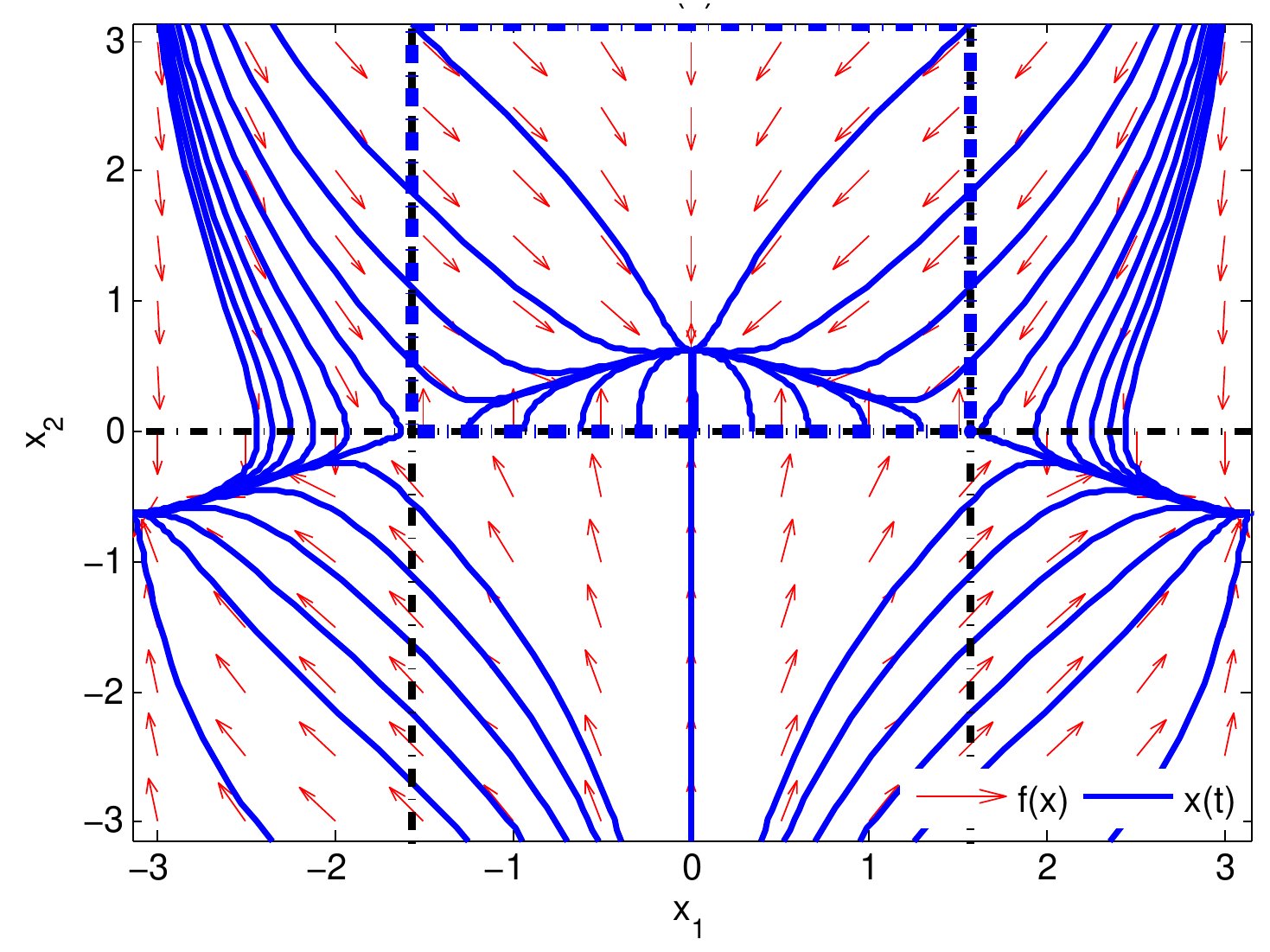}
	\caption{Vector field of (\ref{ex3}) with $a=2, b=2.7, c=1.7$}
	\label{fig4}
\end{figure}

\section{Power system stability}
\subsection{Transient stability}
Although like other nonlinear systems estimating the region of attraction may seem like a reasonable way to determine stability boundaries in a power system.
The power system stability problem can be defined differently due to the requirement of power operators. With protection devices scattered all over power networks with very fast triggering capabilities and redundant connections \cite{kundur}, every fault scenario (in practice) is guaranteed to be cleared, so the interest becomes in what will happen after a fault. In other words, given a fault-on system, if the fault is cleared inside the region of attraction of the post-fault system then the post-fault system is stable and the clearing time is denoted as the critical clearing time which illustrates the potential of accurate estimation of the region of attraction.

%

\subsection{Power system model}

For the purpose of this paper we will consider a classical power system of n synchronous generators with each generator's dynamics represented by the swing equation and
all generators are modeled as constant voltage behind reactance. By assuming fixed impedance loads, the power system model will be governed by the following set of nonlinear
differential equations \cite{bcu}:
\begin{equation} \label{swing}
\begin{aligned}
\dot{\delta}_i    &= \omega_i\\
M_i \dot{\omega}_i &= P_i
-\sum_{j=1}^{n}(V_iV_jB_{ij}\sin(\delta_{ij}))-D_i\omega_i
\end{aligned}
\end{equation}
Where the subscript $i$ represents the machine number, $M_i,D_i$ are the inertia constant and damping coefficient for the $ i_{th}$ machine, respectively, $P_i$ is
the mechanical power input, $V_i$ is the $ i_{th}$ generator's bus voltage magnitude and $B_{ij}$ represents the line admittance. The model describes the dynamics of
two states: $\delta_i$, the generator's angle and $\omega_i$ the generator's angular speed.
This model is known as the classical power system model and have been used extensively in the power system stability literature. Although this model is considered
as a simplified model, assessing stability for such model was proven to be a difficult task \cite{alberto}.
\\
By taking machine $1$ as reference and introducing new variables $\delta_{i,1} = \delta_i-\delta_1, \forall i = [2,n]$, the electric power can be expressed in terms of the new variables as follows:
\[ P_{e,i}(\delta_{2,1},...,\delta_{n,1}) = \sum_{j=1,j\neq 1}^{n} V_iV_jB_{ij}\sin(\delta_{i,1}-\delta_{j,1}) \]

That leads to the following modification to model (\ref{swing}):

\begin{equation} \label{swingm}
\begin{aligned}
\dot{\delta}_{j,1}    &= \omega_j-\omega_1\quad &\forall j = [2,n]\\
M_i \dot{\omega}_i &=
P_i-P_{e,2}(\delta_{2,1},...,\delta_{n,1})-D_i\omega_i\quad &\forall i = [1,n]\\
\end{aligned}
\end{equation}

Which gives the following reduced order model:

\begin{equation} \label{swingred}
\dot{\delta}_{j,1}    = P_j-P_{e,2}(\delta_{2,1},...,\delta_{n,1}) \quad \forall j = [2,n]\\
\end{equation}

The reduced order model (\ref{swingred}) will be used to estimate the critical clearing time using the individual invariance theorem in multi-machine power systems.

\section{Polytopic Approximation method}
In section IV, a region of attraction is estimated as the intersection of polyhedrons that can be described as:
\begin{equation}
\mathcal{P} = \{x\in\Re^n: Ax \leq b\}
\end{equation}

where $A$ is a constant matrix of dimension $m \times n$ and $b$ is a vector in $\Re^m$.\\
As described in the previous section, suppose that a fault-on trajectory is given and denoted as $x^f(t)$. Individual invariance
theorems can be applied algorithmically in a power system transient stability by solving a feasibility problem to make sure that $\mathcal{P}$ is
nonempty and then by substituting $x^f(t)$ in $\mathcal{P}$ to find the first instance at which $\{ x^f(t)\} \bigcap\mathcal{P}=\emptyset$ as defined
more explicitly in the following algorithm:

\begin{algorithm}
	\SetKwInOut{Input}{Input}
	\SetKwInOut{Output}{Output}
	
	\underline{Start}\\
	\Input{$\mathcal{P}$,  $x^f(t)$, $\Delta t$, and $ t_{max}$}
	\Output{$t_c$}
	\eIf{$\exists x : \mathcal{P}(x) \neq \emptyset$}
	{
		go to 7
	}
	{
		go to 11
	}
	{$t=0$}\;
	\textbf{while} {$\{x^f(t)\} \bigcap\mathcal{P}=\emptyset$} \textbf{or} $t = t_{max}$\\
	{$\quad\quad t = t + \Delta t$}\\
	{$\quad\quad t_c = t$}\\
	\textbf{end}
	\caption{Algorithm for finding the critical clearing time given a candidate $\mathcal{P}$ and $\{x^f(t)\}$ }
\end{algorithm}

where $\mathcal{P}$ is the candidate region of attraction and  $x^f(t)$ is the fault-on trajectory. $\Delta t$ is the simulation time step for the fault-on trajectory and $t_{max}$ is the maximum simulation time. In step 2, $\mathcal{P}$ has to be non-empty in order to proceed which can be checked easily by using interior point methods or linear solvers. The algorithm terminates if $\mathcal{P}$ was found to be empty, otherwise, a sequence of function evaluations is executed to maximize $t_c$ as in steps 8-10.

\section{Evaluation and Simulation Results}

Individual invariance theorem can be applied in many cases by defining the individual vectors as follows:

\begin{equation} \label{app1}
\begin{aligned}
\dot{x} =
\begin{bmatrix}
f_1(x) \\
f_2(x) \\
\vdots \\
f_n(x)
\end{bmatrix}
=
\underbrace{\begin{bmatrix}
	f_1(x) \\
	0 \\
	\vdots \\
	0
	\end{bmatrix}}_{f^1(x)}
+
\underbrace{\begin{bmatrix}
	0 \\
	f_2(x) \\
	\vdots \\
	0
	\end{bmatrix}}_{f^2(x)}
+
\dots
+
\underbrace{\begin{bmatrix}
	0 \\
	0 \\
	\vdots \\
	f_n(x)
	\end{bmatrix}}_{f^n(x)}
\end{aligned}
\end{equation}

where each $f^i(x)$ represents a nonlinear interaction with known invariant set such as $\sin(.)$ or $\cos(.)$.\\
A more explicit application of the theorem will be if there are sums of different nonlinear functions, in which case, each function can be separated (if necessary) to construct an individual artificial system as follows:

\begin{equation} \label{app2}
\begin{aligned}
\dot{x}
=
\underbrace{\begin{bmatrix}
	f_1(x) \\
	0 \\
	\vdots \\
	0
	\end{bmatrix}}_{f^1(x)}
+
\dots
+
\underbrace{\begin{bmatrix}
	0 \\
	0 \\
	\vdots \\
	f_n(x)
	\end{bmatrix}}_{f^n(x)}
+
\underbrace{\begin{bmatrix}
	g_1(x) \\
	0 \\
	\vdots \\
	0
	\end{bmatrix}}_{g^1(x)}
+
\dots
+
\underbrace{\begin{bmatrix}
	0 \\
	0 \\
	\vdots \\
	g_n(x)
	\end{bmatrix}}_{g^n(x)}
\end{aligned}
\end{equation}

\subsection{Three-machine 9-bus system}
\begin{figure}
	\centering
	\includegraphics[clip=true,scale=.8]{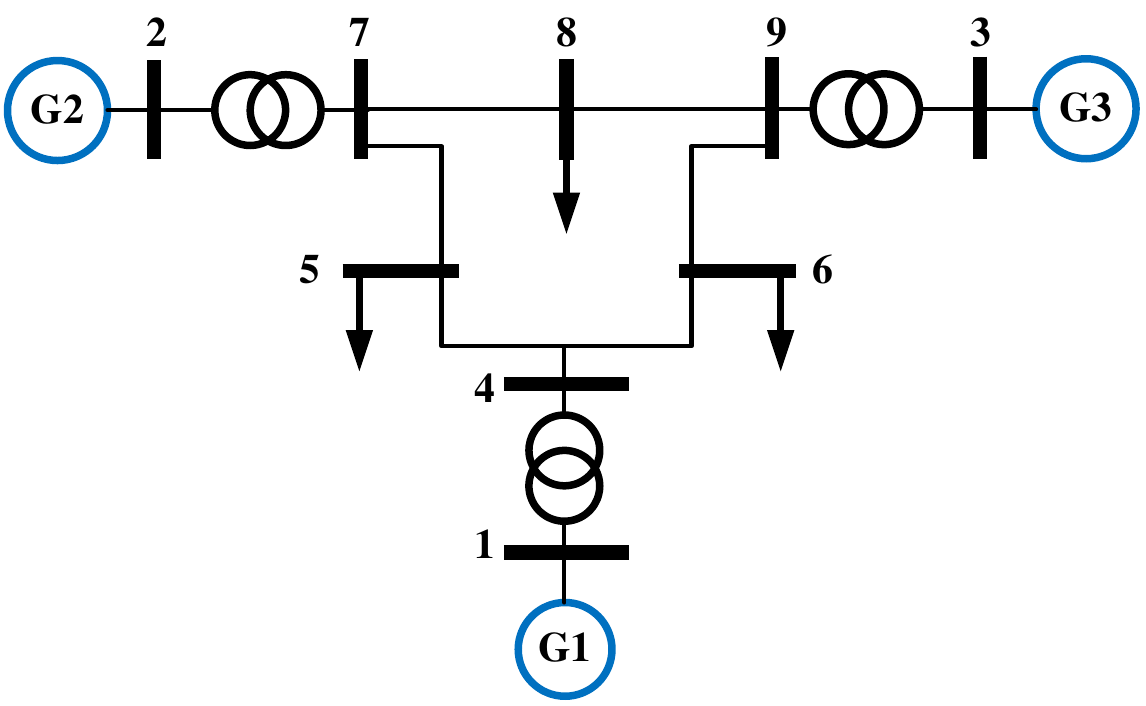}
	\caption{A 3-machine 9-bus system}
	\label{3m9b}
\end{figure}

The three-machine 9-bus system is a network reduced model that was used extensively in power system transient stability literature \cite{chiang}. As depicted in Fig. \ref{3m9b}, the system can be further reduced by reconstructing the Admittance matrix retaining only machines'
buses in a common procedure as given in \cite{chiang,paienergy}. For the required simulations, bus 1 is considered as a slack bus (see Appendix A) which puts the system in the form given in (\ref{swingred}).\\
By appropriate shift of coordinates, each $f^i(x)$ in (\ref{app1}) can be defined as $f^i(x) = P_{e,i}(\delta_{2,1},\dots,\delta_{n,1})$ and is invariant in the set $\Omega_i=\{\delta_{.,1}\in \Re^n: |\sin(\delta_{i,1}-\delta_{j,1})| \leq \pi, \forall i,j\in[2,n]  \} $. In the case of three-machine system, $n=3$, whereas for the 39-bus $n=39$. \\

\begin{table}
	\centering
	\caption{CCT assessment results for 3-machine 9-bus system}
	\label{3m9bT}
	\begin{tabular}{c c c c}
		\multicolumn{1}{l}{}              & \multicolumn{1}{l}{} & \multicolumn{2}{c}{Critical Clearing Time (CCT)} \\ \hline\hline
		\multicolumn{1}{c}{Faulted bus} & Tripped line          & Proposed method      & Time domain simulation     \\ \hline\hline
		\multicolumn{1}{c}{8}   & 8-9    & 0.365 s  & 0.377 s  \\
		\multicolumn{1}{c}{4}   & 4-6    & 0.265 s  & 0.300 s  \\
		\multicolumn{1}{c}{7}   & 7-8    & 0.295 s  & 0.310 s  \\ \hline
	\end{tabular}
\end{table}

\begin{figure}
	\centering
	\subfigure[]{\includegraphics[width=.75\linewidth]{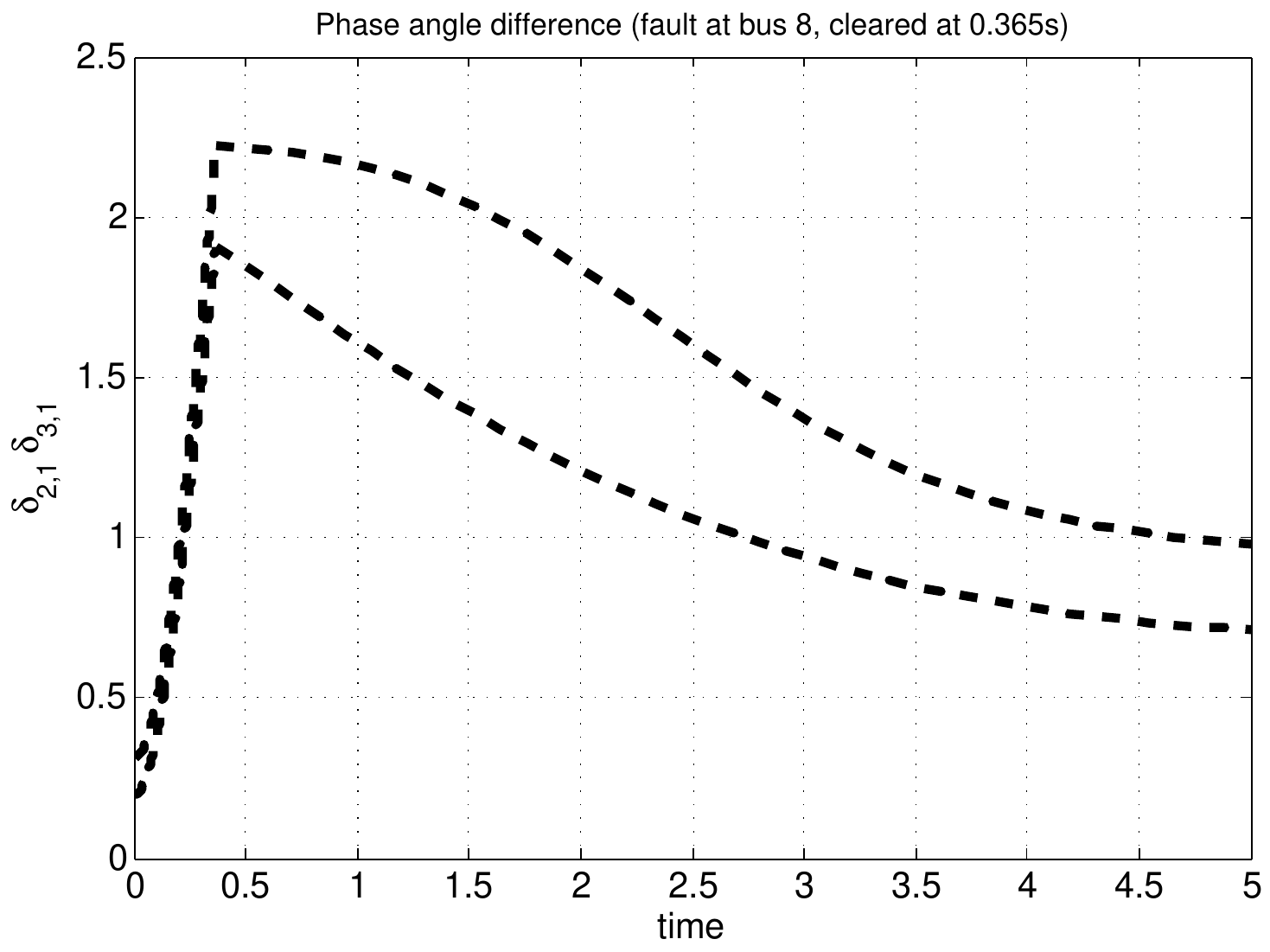}}
	\subfigure[]{\includegraphics[width=.75\linewidth]{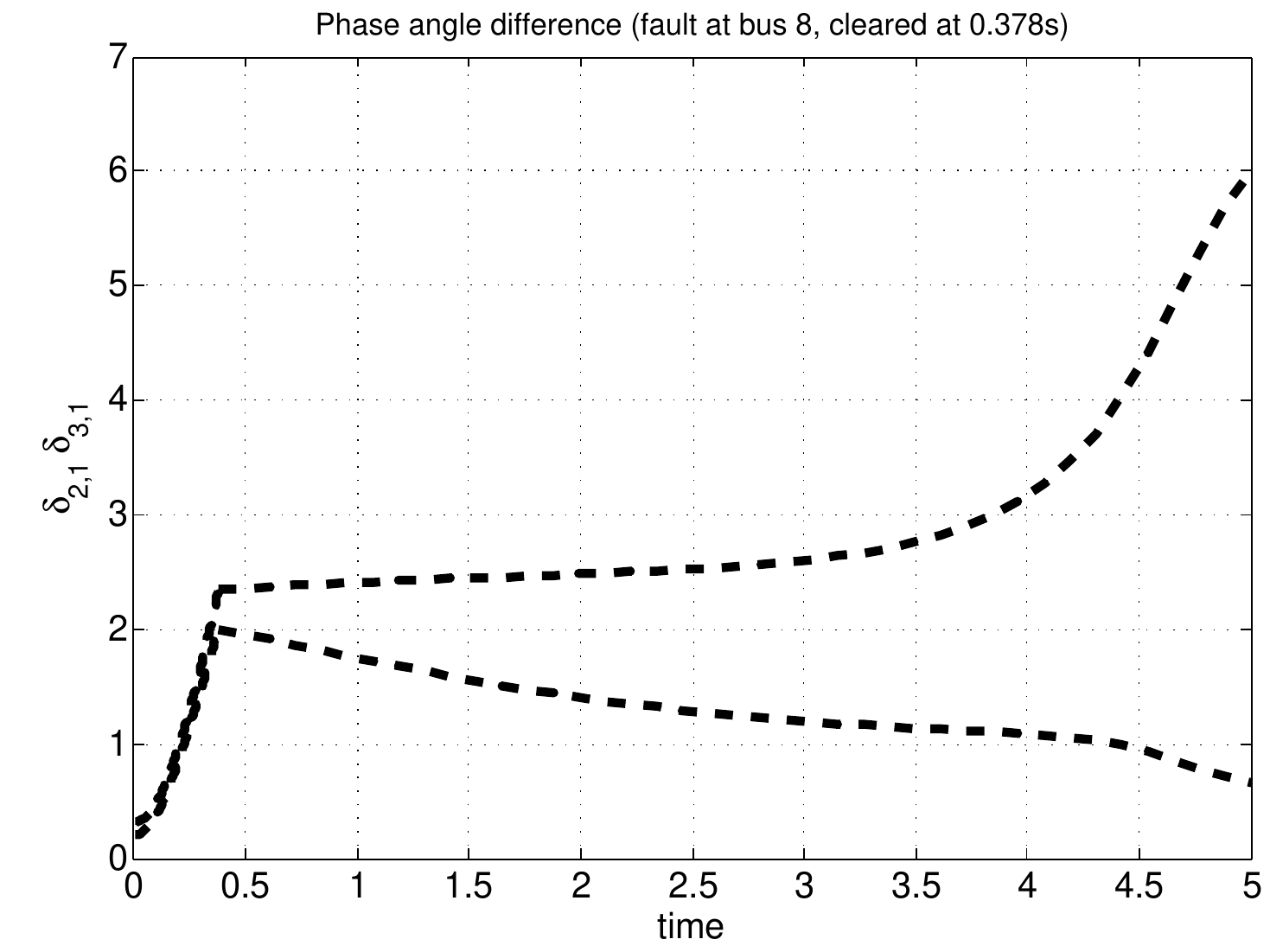}}
	\caption{Time domain results for a fault at bus 8}
	\label{3m9b_8}
\end{figure}

\begin{figure}
	\centering
	\subfigure[]{\includegraphics[width=.75\linewidth]{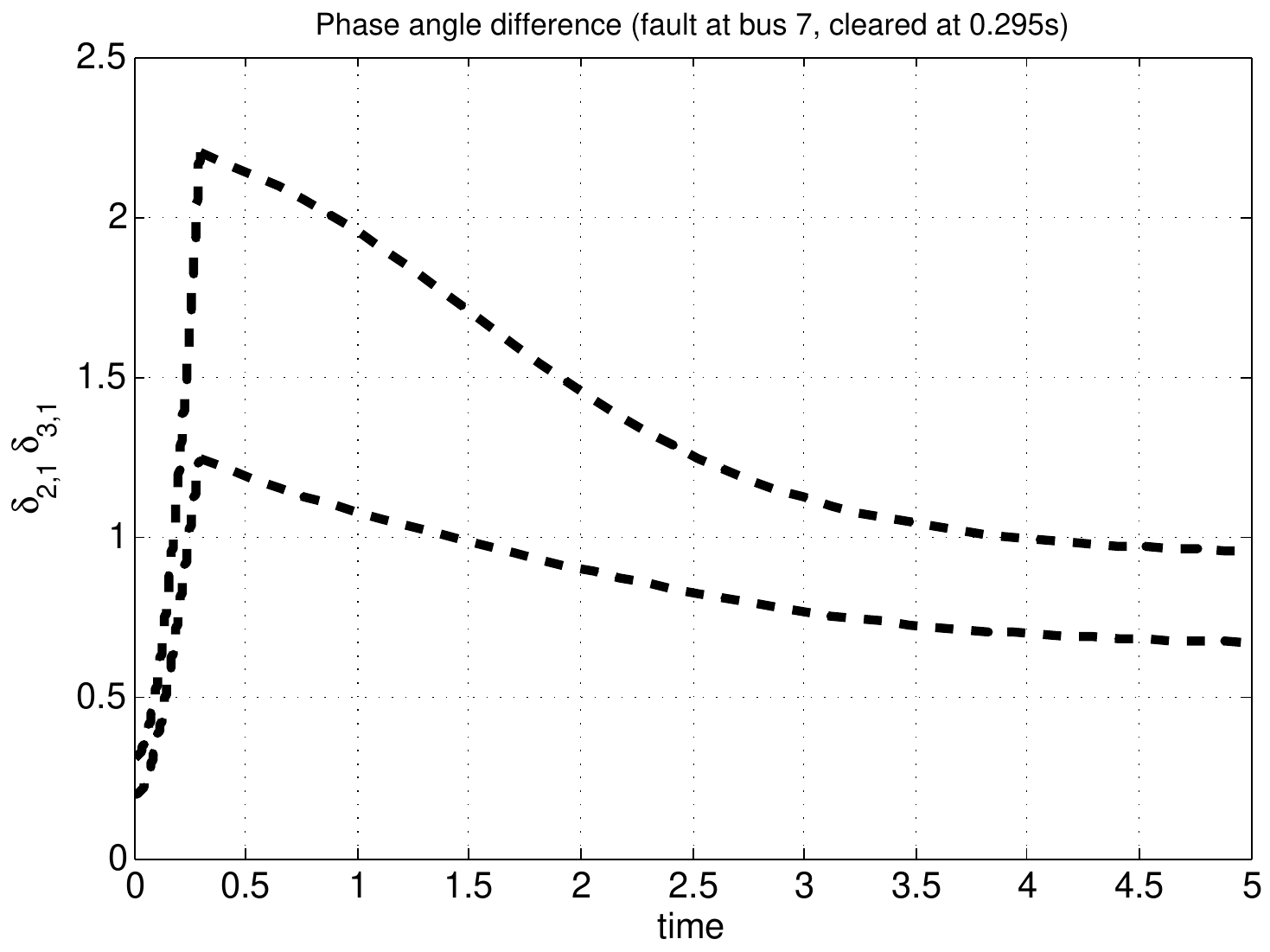}}
	\subfigure[]{\includegraphics[width=.75\linewidth]{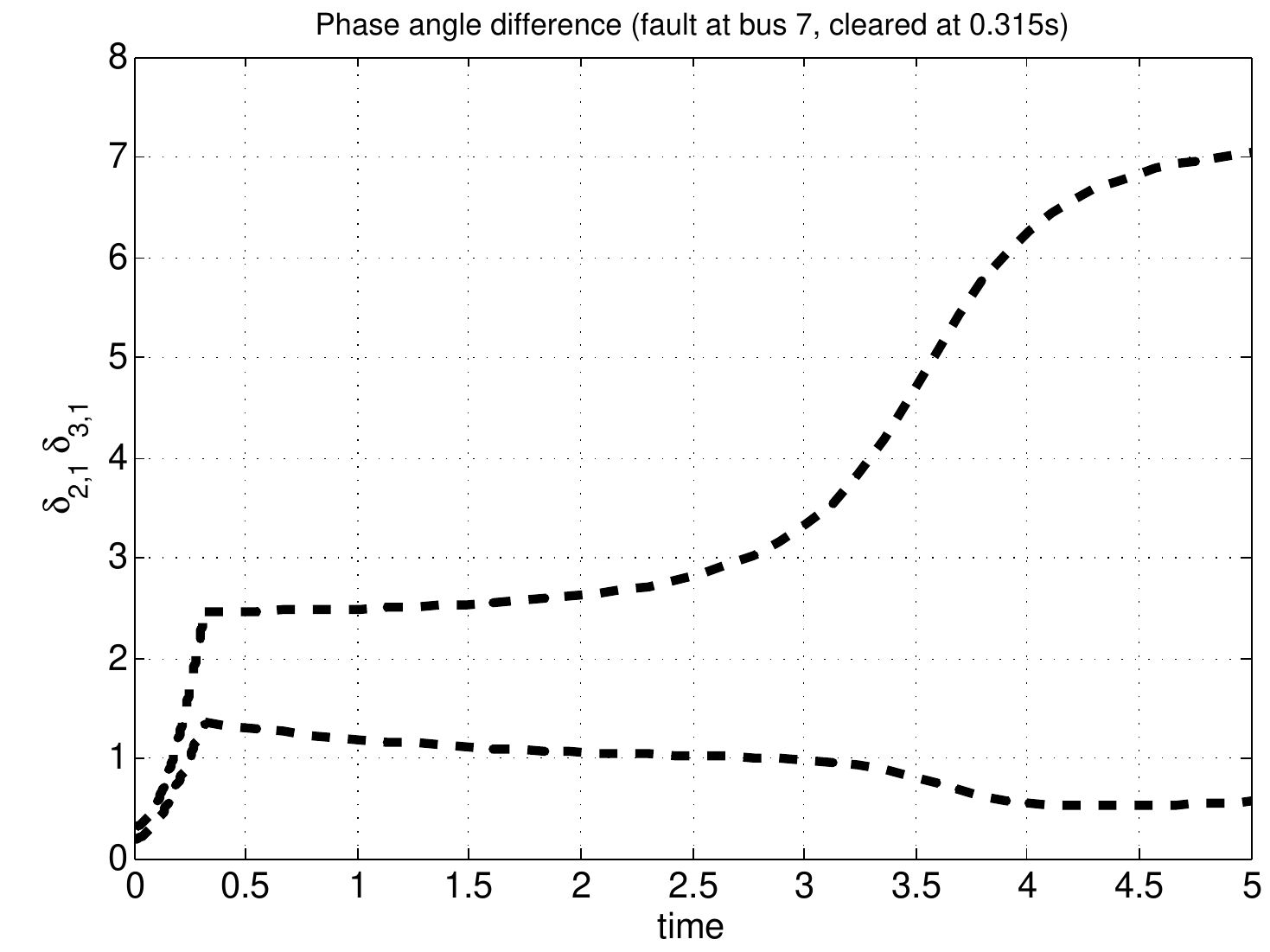}}
	\caption{Time domain results for a fault at bus 7}
	\label{3m9b_7}
\end{figure}

To test the proposed algorithm, a three phase fault is initiated at $t=0$ in the locations given in Table \ref{3m9bT} and is cleared by tripping the associated line. The fault-on trajectory
is calculated by time domain simulation and fed to Algorithm 1 together with the region of attraction estimate $\Omega_e$ in order to find the time at which the fault-on trajectory exits the polytope $\Omega_e$.
It can be seen from Table \ref{3m9bT} that the proposed algorithm gives very close estimation of the critical clearing time with a maximum of 35ms error deviation for the second contingency and a minimum of 13ms for the first contingency. It is noted from Table \ref{3m9bT} that CCT estimates were always to the conservative side and that is expected as the theory provides only sufficient conditions for stability, which is also an advantage given that some algorithms in the literature can lead to overestimates where stability is certified during instable cases (see PEBS method in \cite{chiang}).  Figures \ref{3m9b_8} and \ref{3m9b_7} depict stable cases for the CCT provided by the algorithm and the instable
cases found by exceeding the CCT from time domain simulations which lead to loss of synchronism.

\subsection{New England 39-bus system}
The three machine 9-bus system is a network reduced model that was used extensively in power system transient stability literature. System data are given in Appendix A. Similar to the previous case, Bus 1 was chosen to be the slack bus and the system maintained the same form given in (\ref{swingred}). This case shows a scaled version of the 3-machine test case. Thus, each $\Omega_i$ is found following the same process in the previous section.

\begin{figure}
	\centering
	\includegraphics[clip=true,scale=.4]{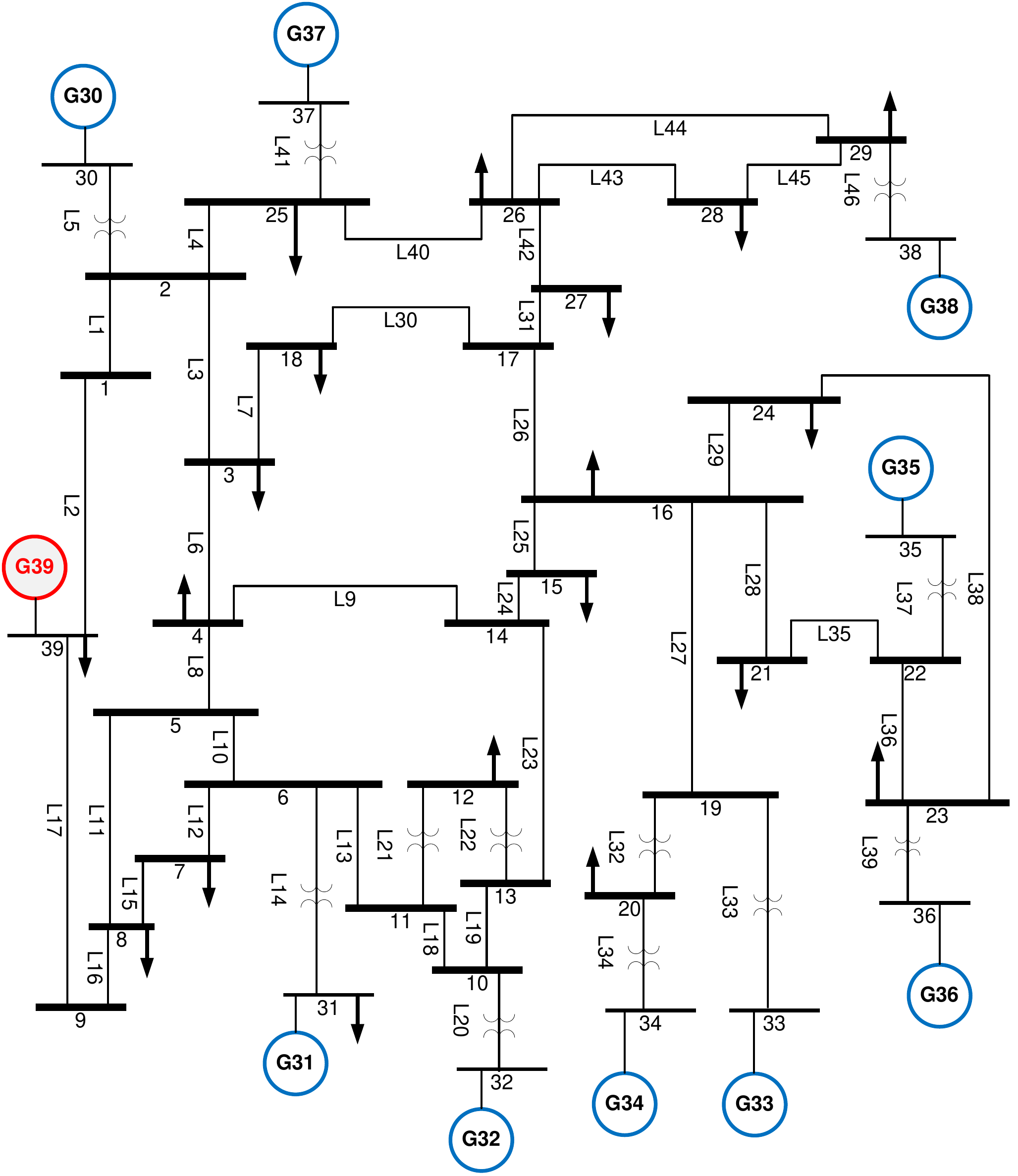}
	\caption{The 39-bus system}
	\label{39b}
\end{figure}

\begin{table}
	\centering
	\caption{CCT assessment results for the 39-bus system}
	\label{39bT}
	\begin{tabular}{c c c c}
	\multicolumn{1}{l}{}              & \multicolumn{1}{l}{} & \multicolumn{2}{c}{Critical Clearing Time (CCT)} \\ \hline\hline
	\multicolumn{1}{c}{Faulted bus} & Tripped line & Proposed method & Time domain simulation     \\ \hline\hline
	\multicolumn{1}{c}{16}   & 16-17    & 0.483 s  & 0.640 s  \\
	\multicolumn{1}{c}{10}   & 10-11    & 0.482 s  & 0.560 s  \\
	\multicolumn{1}{c}{25}   & 25-26    & 0.401 s  & 0.470 s  \\
	\multicolumn{1}{c}{22}   & 22-23    & 0.450 s  & 0.530 s  \\
	\multicolumn{1}{c}{2}    & 2-3      & 0.439 s  & 0.560 s  \\
	\multicolumn{1}{c}{6}    & 6-11     & 0.514 s  & 0.630 s  \\ \hline
\end{tabular}
\end{table}

\begin{figure}
	\centering
	\subfigure[]{\includegraphics[width=.7\linewidth]{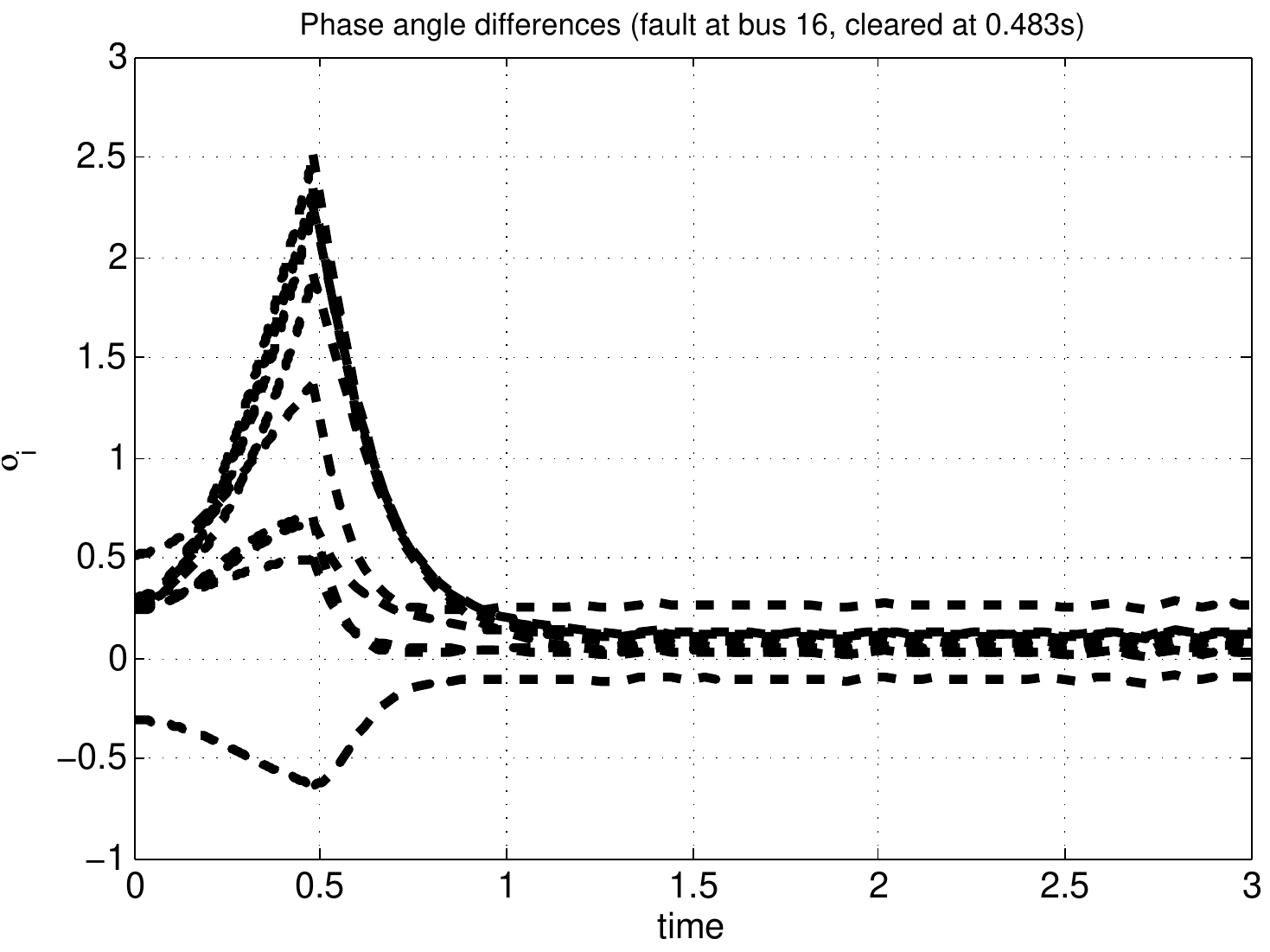}}
	\subfigure[]{\includegraphics[width=.7\linewidth]{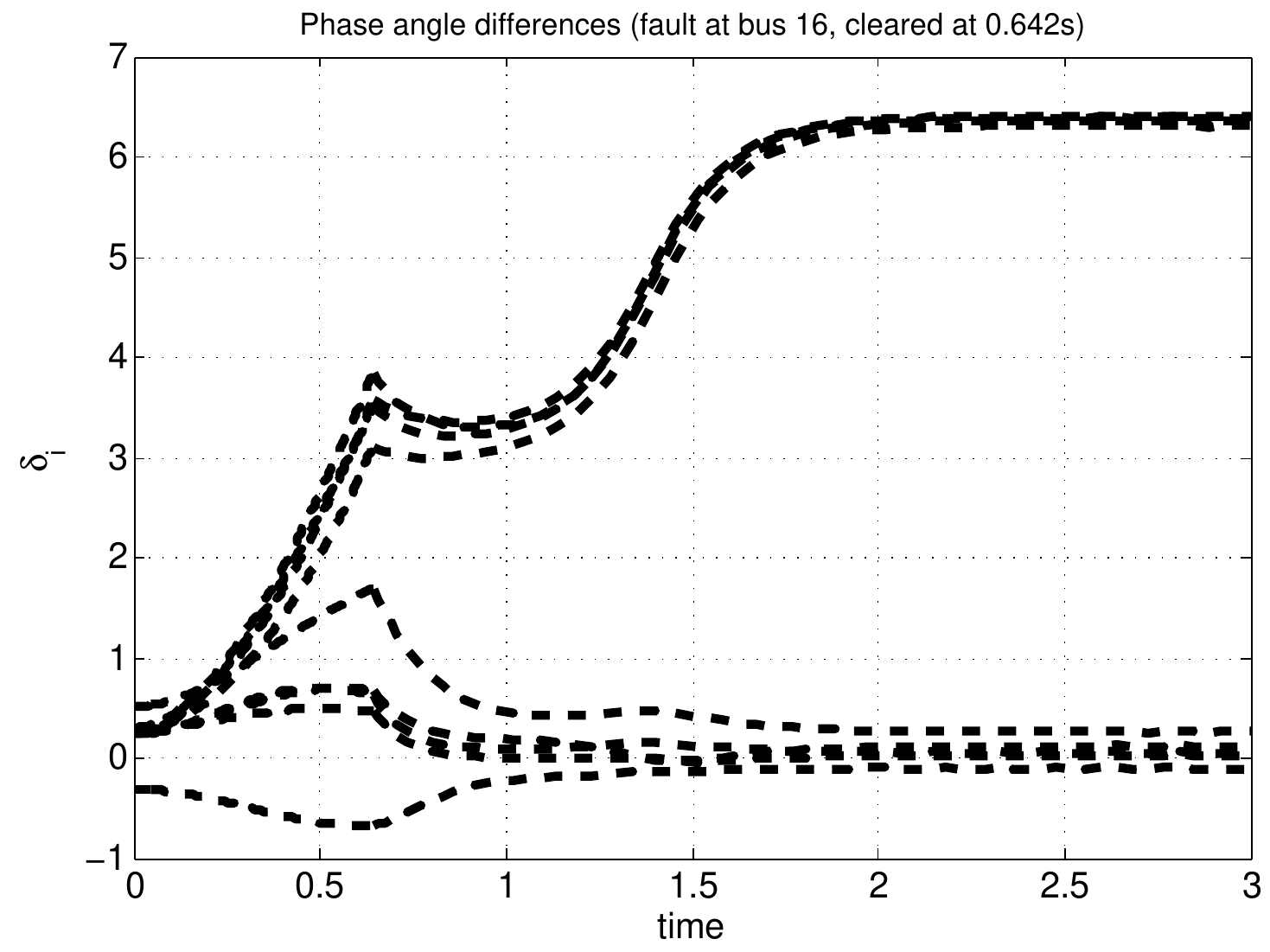}}
	\caption{Time domain results for a fault at bus 16}
	\label{39b_16}
\end{figure}

\begin{figure}
	\centering
	\subfigure[]{\includegraphics[width=.7\linewidth]{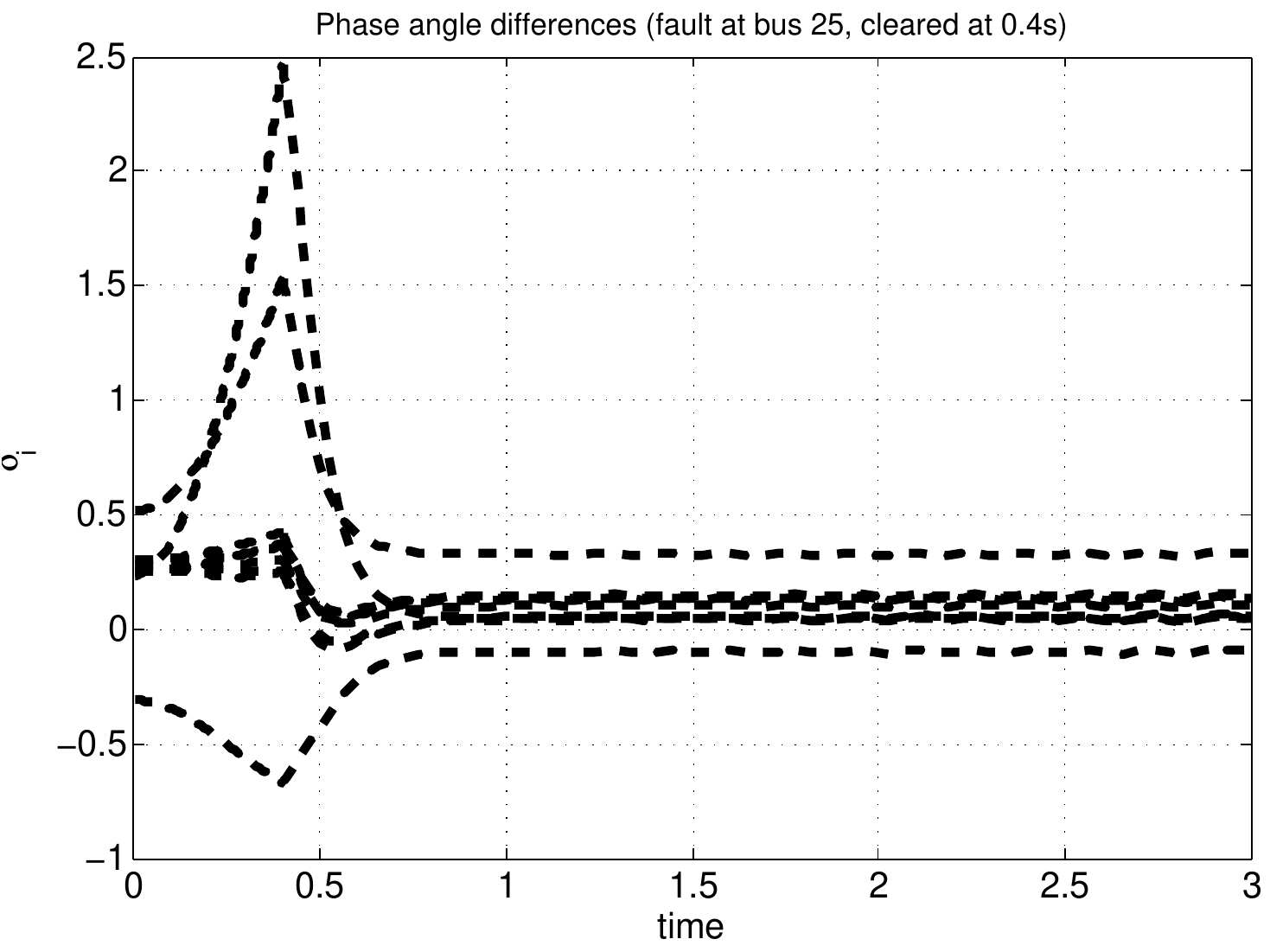}}
	\subfigure[]{\includegraphics[width=.7\linewidth]{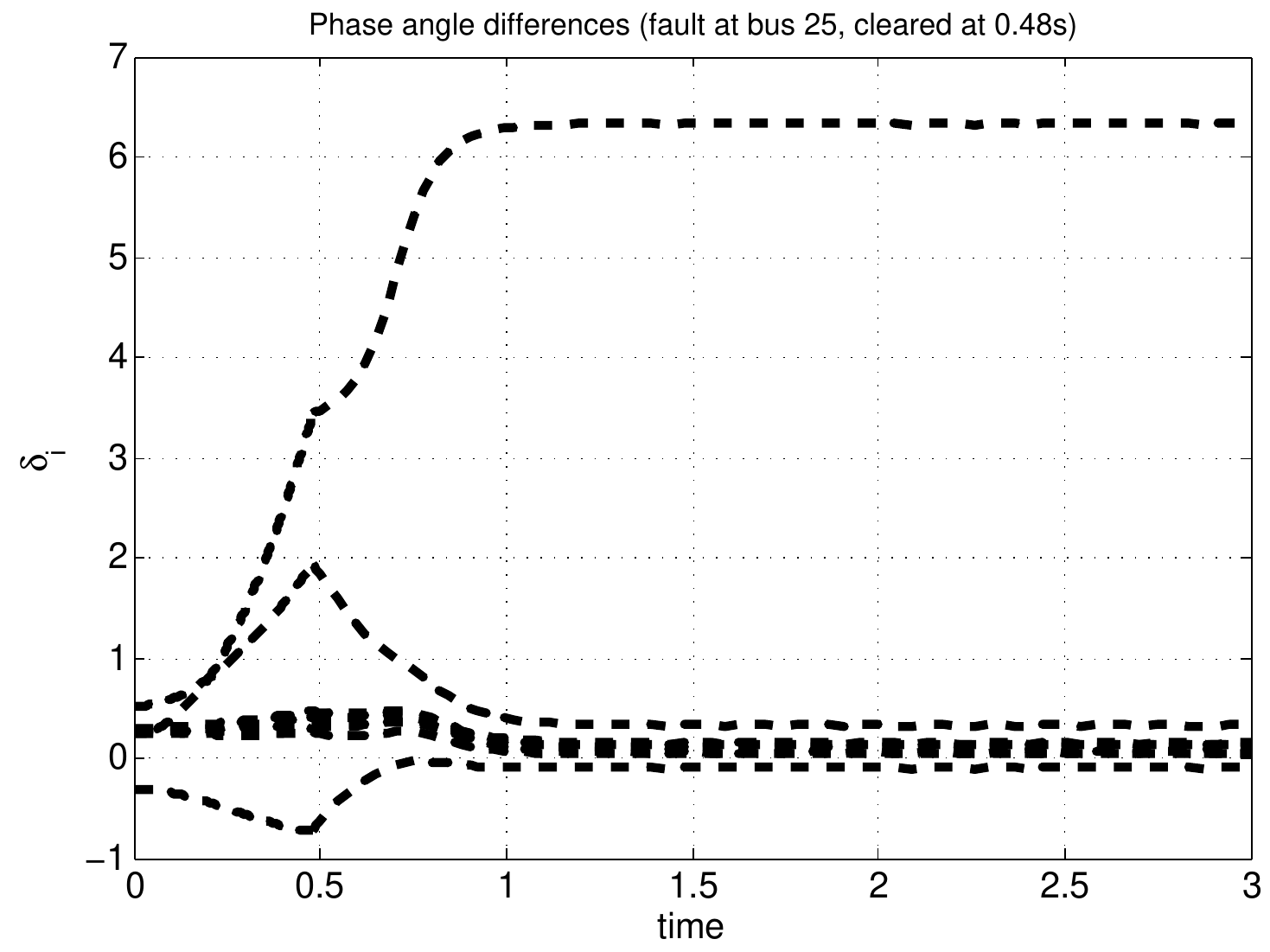}}
	\caption{Time domain results for a fault at bus 25}
	\label{39b_25}
\end{figure}

The goal of this case study is to show the scalability of the algorithm and to examine whether it can provide comparable estimates to time domain simulations. \\
A three phase fault is initiated at $t=0$ in the locations given in Table \ref{39bT} and is cleared by tripping the associated line. The fault-on trajectory is calculated by time domain simulation and fed to Algorithm 1 together with the region of attraction estimate $\Omega_e$ in order to find the time at which the fault-on trajectory exits the polytope $\Omega_e$.\\
It can be seen from Table \ref{39bT} that the algorithm succeeded in providing a very practical estimates of CCT. The maximum deviation occurred for a fault at bus 16 with 157ms whereas the minimum deviation was 69ms for
the contingency at bus 25. Figures \ref{39b_16} and \ref{39b_25} depict the angle deviations for both contingencies, respectively.

\section{Conclusion}
This paper presents a novel algorithm for estimating the region of attraction in nonlinear autonomous systems based on new theoretical findings. A nonlinear vector field is represented as a summation of
individual vector fields of the same dimension, each individual vector field creates an artificial system where its region of attraction is defined independently of the original system. Individual invariance Theorem states that if all individual
vector fields are invariant in the same set then that set is invariant for the original system. This paper provides the theoretical foundation of this novel result
and extends it for an algorithmic construction of regions of attraction. A major advantage of individual invariance theorem is that it allows the use of linear programs
to approximate the region of attraction as an intersection of polyhedrons which provide a computationally efficient approach for transient stability studies
in power system. The presented theory is introduced through several second order examples followed by more practical and high order test cases in power systems.

\appendices
\section{Test systems' parameters}
\begin{center}
\begin{tabular}{c c}
	\hline\hline
	\multicolumn{2}{c}{3-machines 9-bus system \cite{chiang}} \\ \hline\hline	
	\multicolumn{1}{c}{Parameter} 	  & Value 		\\ \hline
	\multicolumn{1}{c}{$H_1$}         & 23.64 s   	\\
	\multicolumn{1}{c}{$H_2$}         & 6.4 s   	\\
	\multicolumn{1}{c}{$H_3$}         & 3.01 s   	\\
	\multicolumn{1}{c}{$D_i$}         & 0.1$H_i$   	\\
	\multicolumn{1}{c}{$V_1$}         & 1.050 pu   	\\
	\multicolumn{1}{c}{$V_2$}         & 1.057 pu   	\\
	\multicolumn{1}{c}{$V_3$}         & 1.022 pu   	\\
	\multicolumn{1}{c}{${x^{'}_d}_1$} & 0.0608 pu  	\\
	\multicolumn{1}{c}{${x^{'}_d}_2$} & 0.1198 pu  	\\
	\multicolumn{1}{c}{${x^{'}_d}_3$} & 0.1813 pu   \\
	\multicolumn{1}{c}{Frequency}     & 50 Hz   	\\ \hline\hline	
\end{tabular}
\end{center}
\begin{center}
\begin{tabular}{c c c c}
	\hline\hline
	\multicolumn{4}{c}{New England 39-bus system \cite{paienergy}} \\ \hline\hline	
	\multicolumn{1}{c}{Parameter} & Value & Parameter  & Value \\ \hline
	\multicolumn{1}{c}{$H_1$}         & 42.0 s  & {${x^{'}_d}_1$} &  0.0310 pu  \\
	\multicolumn{1}{c}{$H_2$}         & 30.2 s  & {${x^{'}_d}_2$} &  0.0697 pu  \\
	\multicolumn{1}{c}{$H_3$}         & 35.8 s  & {${x^{'}_d}_3$} &  0.0531 pu  \\
	\multicolumn{1}{c}{$H_4$}         & 28.6 s  & {${x^{'}_d}_4$} &  0.0436 pu  \\
	\multicolumn{1}{c}{$H_5$}         & 26.0 s  & {${x^{'}_d}_5$} &  0.0660 pu  \\ 	
	\multicolumn{1}{c}{$H_6$}         & 34.8 s  & {${x^{'}_d}_6$} &  0.0500 pu  \\
	\multicolumn{1}{c}{$H_7$}         & 26.4 s  & {${x^{'}_d}_7$} &  0.0490 pu  \\
	\multicolumn{1}{c}{$H_8$}         & 24.3 s  & {${x^{'}_d}_8$} &  0.0570 pu  \\
	\multicolumn{1}{c}{$H_9$}         & 34.5 s  & {${x^{'}_d}_9$} &  0.0570 pu  \\
	\multicolumn{1}{c}{$H_{10}$}      & 31.0 s  & {${x^{'}_d}_{10}$} &  0.0457 pu  \\ 	
\end{tabular}
\end{center}

\begin{center}
	\begin{tabular}{c c c c}
	\multicolumn{1}{c}{$D_i$}         & 0.1$H_i$ & Frequency      & 50 Hz \\ \hline\hline	
	\end{tabular}
\end{center}


\bibliographystyle{IEEEtran}

\begin{thebibliography}{10}
\providecommand{\url}[1]{#1}
\csname url@samestyle\endcsname
\providecommand{\newblock}{\relax}
\providecommand{\bibinfo}[2]{#2}
\providecommand{\BIBentrySTDinterwordspacing}{\spaceskip=0pt\relax}
\providecommand{\BIBentryALTinterwordstretchfactor}{4}
\providecommand{\BIBentryALTinterwordspacing}{\spaceskip=\fontdimen2\font plus
\BIBentryALTinterwordstretchfactor\fontdimen3\font minus
  \fontdimen4\font\relax}
\providecommand{\BIBforeignlanguage}[2]{{%
\expandafter\ifx\csname l@#1\endcsname\relax
\typeout{** WARNING: IEEEtran.bst: No hyphenation pattern has been}%
\typeout{** loaded for the language `#1'. Using the pattern for}%
\typeout{** the default language instead.}%
\else
\language=\csname l@#1\endcsname
\fi
#2}}
\providecommand{\BIBdecl}{\relax}
\BIBdecl

\bibitem{khalil}
H.~K. Khalil, ``Noninear systems,'' \emph{Prentice-Hall, New Jersey}, vol.~2,
  no.~5, pp. 5--1, 1996.

\bibitem{lurepower}
D.~J. Hill and C.~N. Chong, ``Lyapunov functions of lur'e-postnikov form for
  structure preserving models of power systems,'' \emph{Automatica}, vol.~25,
  no.~3, pp. 453--460, 1989.

\bibitem{14}
R.~Genesio, M.~Tartaglia, and A.~Vicino, ``On the estimation of asymptotic
  stability regions: State of the art and new proposals,'' \emph{IEEE
  Transactions on automatic control}, vol.~30, no.~8, pp. 747--755, 1985.

\bibitem{boyd}
L.~Vandenberghe and S.~Boyd, ``Semidefinite programming,'' \emph{SIAM review},
  vol.~38, no.~1, pp. 49--95, 1996.

\bibitem{boyd2}
S.~Boyd, L.~El~Ghaoui, E.~Feron, and V.~Balakrishnan, ``Linear matrix
  inequalities in system and control theory,'' 1994.

\bibitem{parrilo}
P.~A. Parrilo, ``Structured semidefinite programs and semialgebraic geometry
  methods in robustness and optimization,'' Ph.D. dissertation, California
  Institute of Technology, 2000.

\bibitem{kostya}
T.~L. Vu and K.~Turitsyn, ``Lyapunov functions family approach to transient
  stability assessment,'' \emph{IEEE Transactions on Power Systems}, vol.~31,
  no.~2, pp. 1269--1277, 2016.

\bibitem{nonmotonic}
A.~A. Ahmadi and P.~A. Parrilo, ``Non-monotonic lyapunov functions for
  stability of discrete time nonlinear and switched systems,'' in
  \emph{Decision and Control, 2008. CDC 2008. 47th IEEE Conference on}.\hskip
  1em plus 0.5em minus 0.4em\relax IEEE, 2008, pp. 614--621.

\bibitem{bellman}
R.~Bellman, ``Vector lyanpunov functions,'' \emph{Journal of the Society for
  Industrial and Applied Mathematics, Series A: Control}, vol.~1, no.~1, pp.
  32--34, 1962.

\bibitem{wassim}
W.~M. Haddad and V.~Chellaboina, \emph{Nonlinear dynamical systems and control:
  a Lyapunov-based approach}.\hskip 1em plus 0.5em minus 0.4em\relax Princeton
  University Press, 2008.

\bibitem{wassim1}
S.~G. Nersesov and W.~M. Haddad, ``On the stability and control of nonlinear
  dynamical systems via vector lyapunov functions,'' \emph{IEEE Transactions on
  Automatic Control}, vol.~51, no.~2, pp. 203--215, 2006.

\bibitem{wassim2}
S.~G. Nersesov, W.~M. Haddad, and Q.~Hui, ``Finite-time stabilization of
  nonlinear dynamical systems via control vector lyapunov functions,''
  \emph{Journal of the Franklin Institute}, vol. 345, no.~7, pp. 819--837,
  2008.

\bibitem{zbv}
F.~Prabhakara, A.~El-Abiad, and A.~Koivo, ``Application of generalized zubov's
  method to power system stability,'' \emph{International Journal of Control},
  vol.~20, no.~2, pp. 203--212, 1974.

\bibitem{chang1995direct}
H.-D. Chang, C.-C. Chu, and G.~Cauley, ``Direct stability analysis of electric
  power systems using energy functions: theory, applications, and
  perspective,'' \emph{Proceedings of the IEEE}, vol.~83, no.~11, pp.
  1497--1529, 1995.

\bibitem{paienergy}
A.~Pai, \emph{Energy function analysis for power system stability}.\hskip 1em
  plus 0.5em minus 0.4em\relax Springer Science \& Business Media, 2012.

\bibitem{bcu}
H.-D. Chiang, F.~F. Wu, and P.~P. Varaiya, ``A bcu method for direct analysis
  of power system transient stability,'' \emph{IEEE Transactions on Power
  Systems}, vol.~9, no.~3, pp. 1194--1208, 1994.

\bibitem{thorp}
A.~Llamas, J.~D. L.~R. Lopez, L.~Mili, A.~Phadke, and J.~Thorp,
  ``Clarifications of the bcu method for transient stability analysis,''
  \emph{IEEE Transactions on Power Systems}, vol.~10, no.~1, pp. 210--219,
  1995.

\bibitem{6}
C.-C. Chu and H.-D. Chiang, ``Boundary properties of the bcu method for power
  system transient stability assessment,'' in \emph{Circuits and Systems
  (ISCAS), Proceedings of 2010 IEEE International Symposium on}.\hskip 1em plus
  0.5em minus 0.4em\relax IEEE, 2010, pp. 3453--3456.

\bibitem{paganini}
F.~Paganini and B.~C. Lesieutre, ``Generic properties, one-parameter
  deformations, and the bcu method [power system stability],'' \emph{IEEE
  Transactions On Circuits And Systems I: Fundamental Theory And Applications},
  vol.~46, no.~6, pp. 760--763, 1999.

\bibitem{milano}
M.~Anghel, F.~Milano, and A.~Papachristodoulou, ``Algorithmic construction of
  lyapunov functions for power system stability analysis,'' \emph{IEEE
  Transactions on Circuits and Systems I: Regular Papers}, vol.~60, no.~9, pp.
  2533--2546, 2013.

\bibitem{sos}
G.~Chesi, \emph{Domain of attraction: analysis and control via SOS
  programming}.\hskip 1em plus 0.5em minus 0.4em\relax Springer Science \&
  Business Media, 2011, vol. 415.

\bibitem{Ross}
K.~A. Ross, \emph{Elementary analysis}.\hskip 1em plus 0.5em minus 0.4em\relax
  Springer, 1980.

\bibitem{kundur}
P.~Kundur, J.~Paserba, V.~Ajjarapu, G.~Andersson, A.~Bose, C.~Canizares,
  N.~Hatziargyriou, D.~Hill, A.~Stankovic, C.~Taylor \emph{et~al.},
  ``Definition and classification of power system stability ieee/cigre joint
  task force on stability terms and definitions,'' \emph{IEEE transactions on
  Power Systems}, vol.~19, no.~3, pp. 1387--1401, 2004.

\bibitem{alberto}
N.~G. Bretas and L.~F. Alberto, ``Lyapunov function for power systems with
  transfer conductances: extension of the invariance principle,'' \emph{IEEE
  Transactions on Power Systems}, vol.~18, no.~2, pp. 769--777, 2003.

\bibitem{chiang}
H.-D. Chiang, \emph{Direct Methods for Stability Analysis of Electric Power Systems}.\hskip 1em plus 0.5em minus 0.4em\relax John Wiley \& Sons, Inc.,
  2010.

\end{thebibliography}


\end{document}